# Metabolic robustness and network modularity: A model study

Petter Holme

Department of Physics, Umeå University, 90187 Umeå, Sweden; Department of Energy Science, Sungkyunkwan University, Suwon 440–746 Korea

**Background** Several studies have mentioned network modularity—that a network can easily be decom- posed into subgraphs that are densely connected within and weakly connected between each other—as a factor affecting metabolic robustness. In this paper we measure the relation between network modularity and several aspects of robustness directly in a model system of metabolism.

**Methodology/Principal Findings** By using a model for generating chemical reaction systems where one can tune the network modularity, we find that robustness increases with modularity for changes in the concentrations of metabolites, whereas it decreases with changes in the expression of enzymes. The same modularity scaling is true for the speed of relaxation after the perturbations.

**Conclusions/Significance** Modularity is not a general principle for making metabolism either more or less robust; this question needs to be addressed specifically for different types of perturbations of the system.

## Background

Graph theoretical methods are useful to study the large-scale organization of biological systems (*1*). One such system is the metabolism—the set of chemical reactions needed to sustain the normal, healthy state of an organism. We call a graph derived from a metabolic reaction system a *metabolic network*. One of the main findings from statistical studies of metabolic networks is that the metabolism has larger *network modularity* (*2,3*)—the tendency for a network to be divisible into subgraphs that are densely connected within, and sparsely connected between each other—than expected (*4*). However, metabolic networks are far from perfectly modular—no matter how the network modules are defined, there will be plenty of connections between them (*4–8*). The network modules are often interpreted as biological modules---functionally independent subunits (*9*). This interpretation is a natural consequence of interpreting edges as functional couplings of relatively equal strength. Despite the lack of comprehensive experimental evidence, metabolism is assumed to be robust to e.g. changes in concentration of metabolites (*10*). Modularity is often thought to contribute to the robustness of various biological systems (*11–13*). But if this is true for metabolism too, that modularity contributes to both functionality and robustness, then how come there are so many cross-modular couplings? One explanation could be that these couplings are inevitable—the laws of physics give no way of avoiding intermodular connections. Another explanation could be that the intermodular edges actually stabilize the system so that the organization we observe is a compromise where adding functionality increases modularity and adding robustness decreases modularity. Such a role of modularity relates to the concept of *distributed robustness* (*14*)—if a module fails, many other modules can collectively compensate for this loss, there need not be any replacement module. In terms of metabolic networks, this means that there will be many connections between the modules and thus that the network modularity will be comparatively low. In this paper we investigate the role of network modularity in large chemical reaction systems as directly as possible—by measuring the system's response to different types of perturbations in a model with tunable network modularity.

Our simulations start by generating a chemical reaction system. This generative algorithm is stochastic and by tuning the input parameters, we can control the expected network modularity (Fig. 1) (*15*). Then we generate a random distribution of metabolites and relax the system to equilibrium (using mass-action kinetics with an implicit enzymatic control). From this state, we apply a certain type of perturbation to the system and let it relax to a new equilibrium. To quantify robustness, we measure how close the two equilibria are to each other. We also measure the relaxation time, i.e. how fast the system can respond to the perturbation (and for that reason, we do not employ faster calculations of the equilibrium state (*16,17*)). In Fig. 2 we show an example of these steps. As the reaction system is generated by a stochastic method we repeat the procedure above to obtain averages. For each value of the input parameters, we measure average values over 500 realizations of all steps above of both the network modularity and the quantities characterizing robustness. From these data points we derive trends in the modularity-dependence of different aspects of robustness.

## Results and Discussion

### Robustness as a function of network modularity

Robustness is a broad concept that hardly can be condensed into one measure, even for a system as specific as metabolism. In general, robustness can be defined as a system's ability to remain unchanged when perturbed. One can imagine several types of perturbations. We investigate two rather different classes—changes in concentrations of metabolites and changes in the reaction system (new reactions replacing old) by genetic control. We refer to the first case as *metabolic* perturbations and the second as *genetic* perturbations. We will also distinguish between: if the perturbations are localized to one module, or if they can appear anywhere in the network. In total we consider four classes of perturbations—they can be either localized or global, and metabolic or genetic.

The main robustness measure, defined in the Methods section, is basically the relative change in the concentration of a metabolite averaged over a set of metabolites. We consider two such sets, either the whole set of metabolites, which gives the *system-wide robustness r*, or the metabolites that are perturbed giving the *focal robustness r***.

In Fig. 3, we plot the average values of our robustness measures as functions of the average network modularity $q$. The robustness to global metabolic perturbations increases while the robustness to perturbations within a module remains fairly constant (Fig. 3A). If one looks only at the metabolites that were originally perturbed (Fig. 3A), the situation is different---these metabolites are more affected by sudden shifts in the concentrations the more modular the system is. This seems logical—if the modularity is lower, the coupling to the rest of the network is stronger, so there are more metabolites to influence the relaxation and to absorb the perturbation. The fact that the system is more robust to global, compared to localized, perturbations can be explained in a similar way—a localized perturbation gives a larger impact on a restricted subsystem and this subsystem cannot absorb that large impact as much as the whole system would. But why does the system-wide robustness increase with modularity? One scenario is that metabolic perturbations are better absorbed in a distributed fashion. With global perturbations and high modularity each module handles its internal perturbations and, if this fails, flows between the modules are too weak for the perturbation to spread.

For the genetic perturbations all curves are decreasing, meaning that modularity makes the system less robust. These perturbations virtually add new reactions and delete old. Even if the perturbations are designed not to affect the average structure of the system (keeping e.g. the system size $R$ and the modularity $q$ constant), they obviously affect $r$ more than the metabolic perturbations (cf. Fig. 3A and Fig. 3C). A network module can presumably not handle a genetic perturbation as efficient as a metabolic perturbation. Another factor for the decreasing $q(r)$-curve could be that the interface between the modules can change from the perturbations and that the interfaces get more influential with increasing modularity. As seen in Fig. 3D, the localized perturbations influence the directly affected metabolites (the ones that are involved in reactions changed by the genetic perturbations) less strongly than the global perturbations. From the changes at the interfaces, we can understand that localized perturbations affect the rest of the system to a greater deal here than compared with metabolic perturbations. $r^*$ is larger for the local compared with global genetic perturbations meaning that for metabolites within a single module rewired by genetic perturbations the changes will be larger than if the perturbations are more distributed.

### Relaxation time as a function of network modularity

In Fig. 4, we show the relaxation time τ as a function of modularity. A small τ value means that the system reaches its new equilibrium fast. This dynamic response is different for the two types of perturbations—the system reaches its new state faster with higher modularity for the metabolic perturbations, but slower with the genetic ones. The decreasing $\tau(q)$ curves for metabolic perturbations is in line with the above mentioned scenario that if modules handle the perturbations independently, then the more modular the system is the better (in this case faster) is the recovery. That, for genetic perturbations, robustness increases with modularity is something we interpret as an effect of the changed couplings across at the boundary. The stronger the modularity is, the slower is the flow between the modules and the longer does the system need to find a new equilibrium.

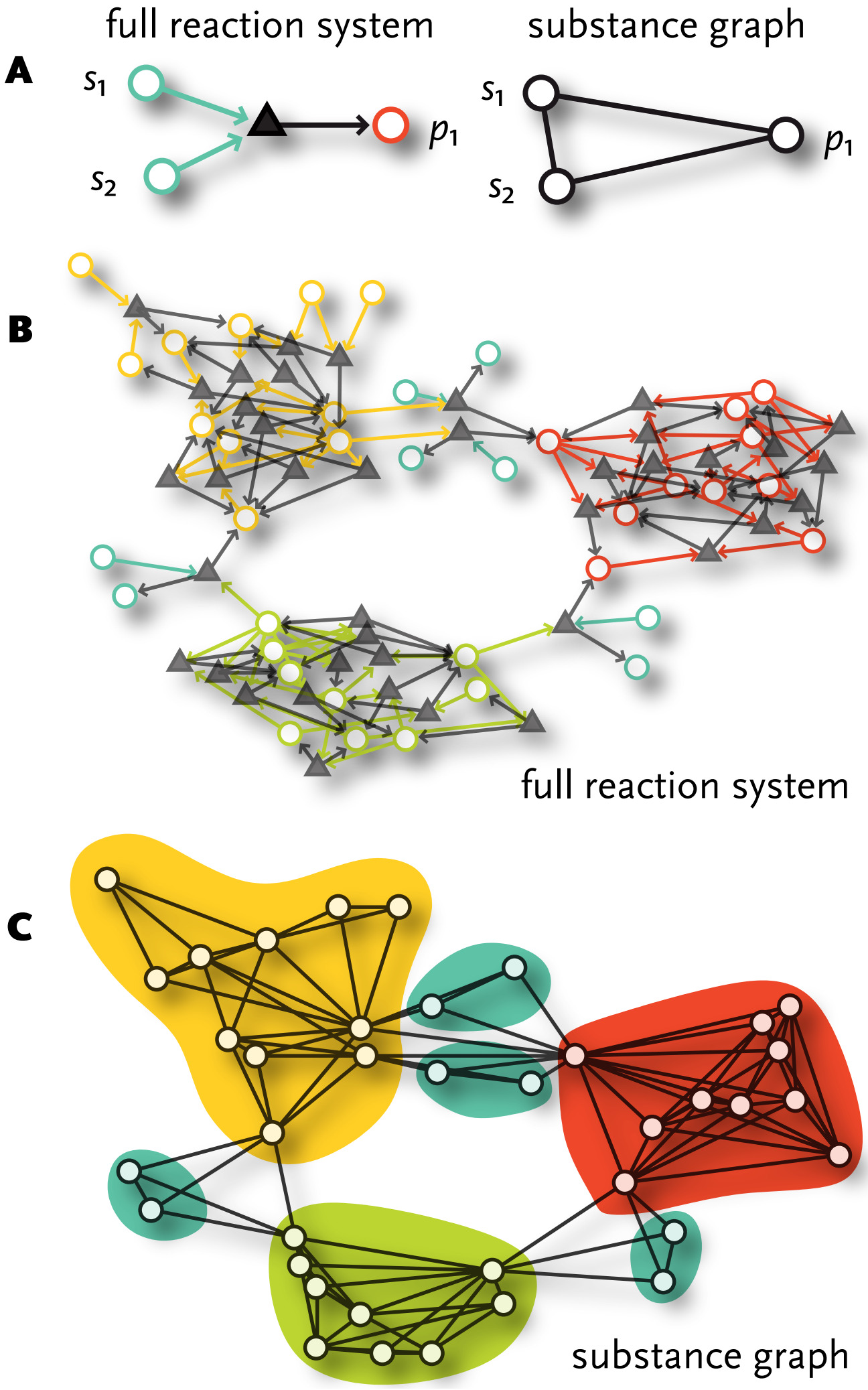

**Fig. 1. Example of the reduction from reaction systems to substance graphs and the generation of modular reaction systems.** In A we see how the two substrates and one product (circles) of a reaction (triangle) gets reduced to a substance graph. An arrow going into a reaction marks the substrate, an arrow going out marks the product. Panel B illustrates a reaction system obtained with the method of the manuscript. The parameter values for this reaction-system example are $R=4$, $g=3$, $n_g=3$, $n_{trial}=100$ and $\gamma=0.9$. The algorithm proceeds by assembling reactions and metabolites in disjoint clusters (the three larger clusters of distinct colors). Then we add a fraction of metabolites and reactions that can connect to any parts of the system. The larger this fraction of global reactions is, the lower is the network modularity of the projected network.

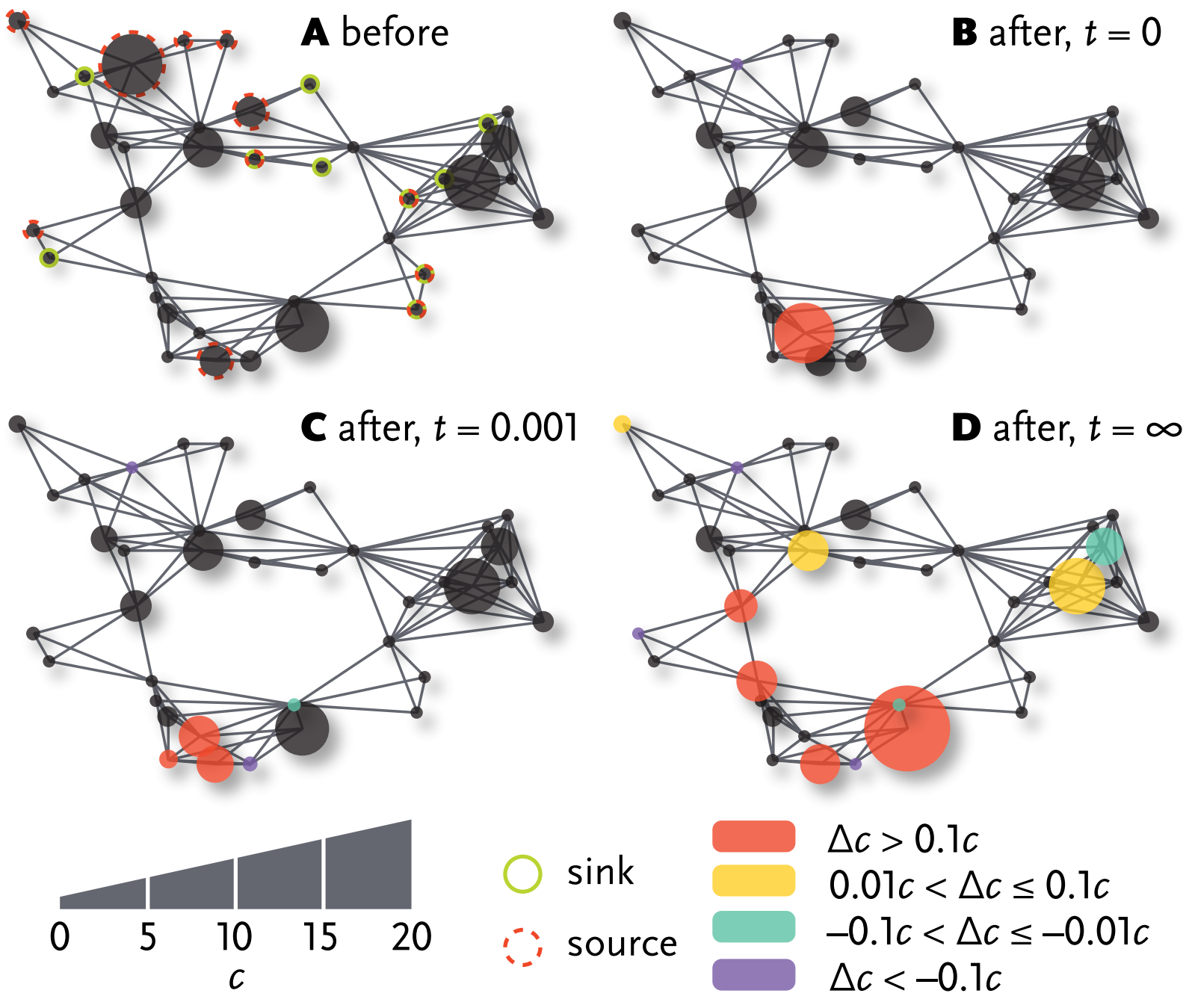

**Fig. 2. The procedure to measure robustness.** The figure illustrates a reaction system at equilibrium visualized by its reaction graph A, getting perturbed by redistributing the mass of (in this case two) metabolites B and how the system relaxes to another equilibrium (C and D). The concentration is illustrated by the size of the circles (the total mass, not the concentration is conserved, so the total areas of the circles are not the same in the different panels). The change in concentration is indicated by color. A metabolite unaffected by the perturbation is colored black.

## Conclusions

We have, in a model framework, directly measured the effects of network modularity on the robustness of chemical reaction systems. The main conclusion is that modularity does affect robustness but not in a unique way. Modularity is thus, it seems, not a general principle for either strengthening or weakening robustness, not even in such a specific system as metabolism. When relating robustness and modularity, one needs to specify what kind of perturbation we measure robustness against. For sudden changes in concentration levels, in our model, more modular reaction systems are more robust and converge to an equilibrium state faster than less modular systems. If, on the other hand, the genetic control is altered—so that other enzymes are expressed—then modularity decreases robustness. In an evolutionary perspective, this essentially means that we need more detailed studies. Real metabolic networks are more modular (in the network-modularity sense) than random networks, but still far from, say, a system engineered by humans (*18*). One scenario is that robustness is key driving force in evolution of metabolic-network structure and that this weakly modular structure above comes from trade-offs between robustness-increasing and robustness-decreasing changes in modularity. However, functionality and chemical constraint probably also play a major role in this evolution. Note that if one considers smaller feedback loops as modules, rather than network clusters, evolution is by necessity modular in the sense that adding the production of a new substance often needs the addition of its degradation (this is because many substances cannot penetrate the cell membrane and would be toxic if accumulated). The conclusion that modularity does not affect robustness in a single direction has further implications for synthetic biology that often, at least theoretically, strives to design functionality from combination of modules (*19*,*20*)—our study hints the such an approach would not give robustness for free.

For the future, we anticipate more studies cataloguing the principles of robustness, and the effects of modularity. We believe model studies like the present are the best theoretical way to proceed. An alternative is to compare the modularity of different organisms (*21*) to find changes in the modularity over the course of evolution, but in such an approach it would be hard to tease apart fundamental physical constraints from evolutionary pressure. It would of course also be interesting to experimentally compare the response of different organisms, or cell types, with metabolism of different network modularity to perturbations. Further into the future, we hope for experimental methods to measure the dynamics of the entire chemical composition of cells.

## Methods

### Notations and mathematical framework

We consider a reaction system of $N$ metabolites **M** and $R$ reactions **R**. A reaction $r \in \mathbf{R}$ is characterized by its substrates $s_1,\ldots,s_{S(r)} \in \mathbf{M}$, their multiplicities $\sigma_1,\ldots,\sigma_{S(r)}$, its products $p_1,\ldots,p_{S(r)} \in \mathbf{M}$, and their multiplicities $\pi_1,\ldots,\pi_{S(r)}$, and a reaction coefficient $\kappa_r$. Consider, for example, the reaction $2H + O_2 \rightarrow 2H_2O$. Then we have $S=2$, $P=1$; $s_1$ is $H_2$, $s_2$ is $O_2$, $\sigma_1=2$, $\sigma_2=1$, $p_1$ is $H_2O$ and $\pi_1=2$. From a reaction system one can derive a graph $G=(V,E)$, where $V$ ($V=\mathbf{M}$ in this case) is the set of vertices of the graph and $E$ is the set of edges. One can define several types of metabolic graphs. In this work we focus on substance graphs (claimed to encode more functional information about the graphs than other simple-graph representations (*5*,*15*)), where the vertices are substances and there is an (undirected) edge between two vertices if they are either substrates or products of the same reaction (edges between a vertex to itself is not allowed). In the example above, the reaction will contribute with three edges—($s_1$, $s_2$), ($s_1$, $p_1$) and ($s_2$, $p_1$)—to the substance graph (see Fig. 1A).

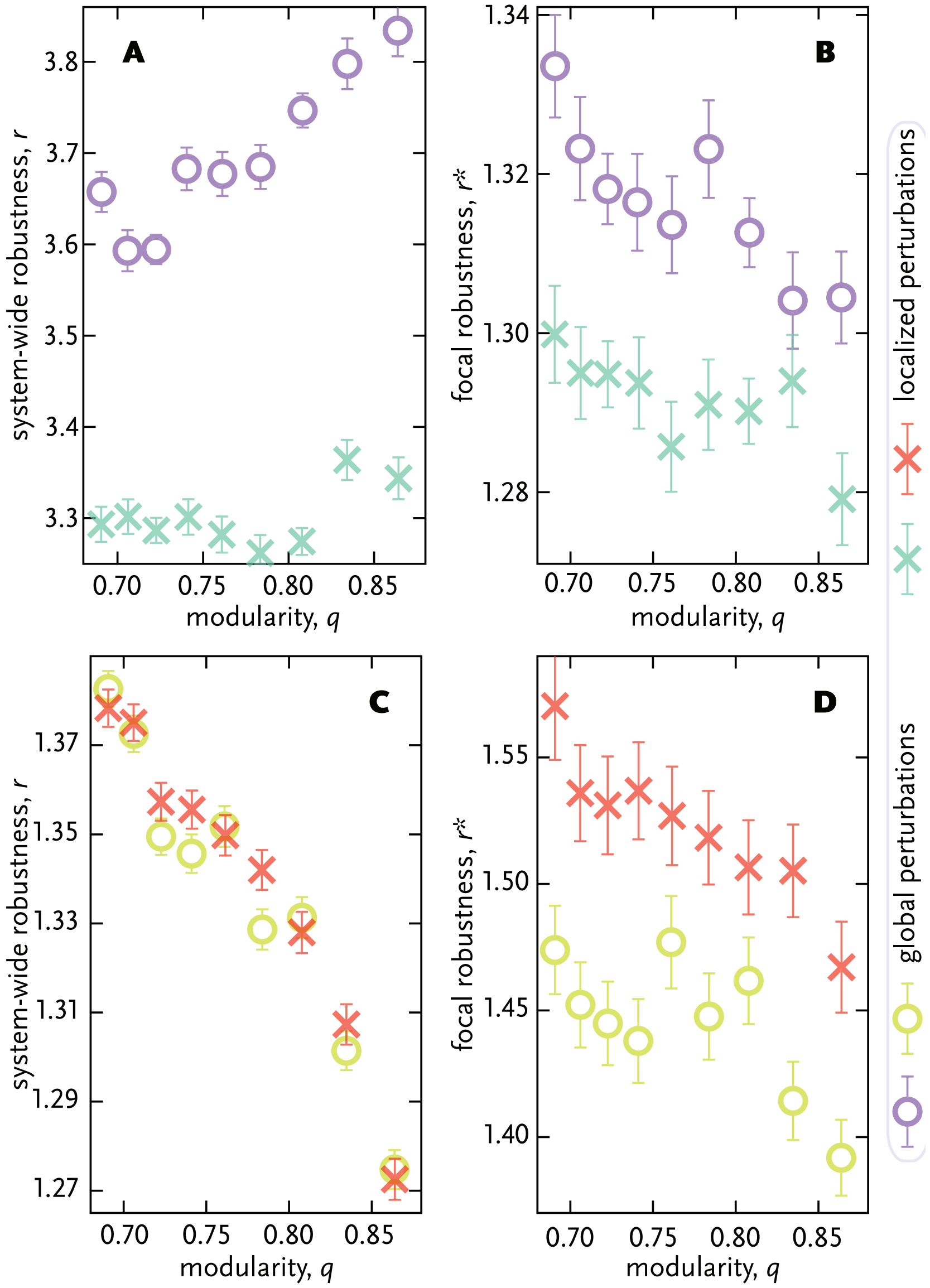

**Fig. 3. Ronustness vs. modularity.** Panels A and B show data for the robustness against metabolic perturbations. A displays robustness of the system as a whole; B shows the robustness measured over the perturbed metabolites only. Panels C and D show the corresponding plots for robustness against genetic perturbations. Circles represent perturbations made in one module; crosses indicate data for perturbations made in different modules. The data is averaged over more than 500 runs (network realizations). The errorbars in the average $q$ are smaller than the symbol size.

### Network modularity

We will shortly discuss how network modularity is calculated. For a more comprehensive review, see Refs. (2,3). Let the vertex set be partitioned into groups and let $e_{ij}$ denote the fraction of edges between group $i$ and $j$. The modularity of this partition is defined as

$$Q = \sum_i \left[ e_{ii} - \left( \sum_j e_{ij} \right)^2 \right] \quad (1)$$

where the sum is over all the vertex groups. The term $(\sum_j e_{ij})^2$ is the expectation value of $e_{ii}$ in a random graph. The measure for graph modularity that we use is $q(G)$—$Q$ maximized over all partitions (by a heuristics proposed in Ref. 3). Comparing $q$ of graphs with different sizes and degree distributions is not completely straightforward. Even for networks generated by one particular model (that one would from construction expect to have the same modularity) $q$ can vary with the network size (22). Fortunately, for this work, such changes are monotonous. This means that we can use $q$ to detect changes in robustness in response to changes in modularity even though the particular functional forms of the curves of robustness vs. $q$ are hard to interpret.

### Model reaction systems with tunable network modularity

In this section, we will sketch the model of reaction systems with tunable network modularity. The model we use treats atoms of the molecular species explicitly. The set of all atoms is divided into $g$ groups (or proto-modules) of equal size $n_g$. $R$ reactions are added to the system such that they obey mass conservation (for all atom species, the number of individuals is the same for substrates and products). $\gamma r$ reactions are added between molecules consisting of atoms from the same group. The remaining $(1-\gamma)r$ reactions are added between molecules of any atomic composition. For low $\gamma$-values, relatively few reactions will connect different groups and therefore the derived network modularity will be low. If $\gamma$ is close to one, the derived graphs will be more modular. The molecules are constructed by randomly combining atoms of a group. Reactions are generated by randomly splitting and recombining molecules. If the mass conservation is broken, or the reaction already exists in the data set, then the molecule construction is repeated. If no reaction fulfilling mass conservation has been found after $n_{\text{trial}}$, iterations then this is done by defining new molecules. With a larger value of $n_{\text{trial}}$, the substance graphs will thus be both denser and have fewer metabolites ($N$ is, perhaps a little unusually, an output of the model, whereas $R$ is a control parameter).

There are a number of other technicalities, like how the molecules are constructed from the atoms etc., that are explained in detail in Ref. (15). We also modify the algorithm of Ref. (15) when it comes to inter-group reactions. In Ref. (15) these always act as sources and sinks (so that there is never a flow between modules); here all inter-group reactions are bridges between the modules (so that these reactions have at least one substrate in one group and one product in the other).

In this work we use the parameter values $R=500$, $g=10$, $n_g=5$ and $n_{\text{trial}}=100$ (the values of the other parameters, related to the details in Ref. (15) are the same as in that paper).

### Reaction kinetics

To simulate the biochemical dynamics, we use simple mass-action kinetics. This approach is, technically speaking, assum-ing all enzymatic effects can be encoded into the reaction coefficients and the reaction system itself. The main reason for this simplification is that, when speaking about network modularity, enzymes are usually only included implicitly (via the active reactions), so to relate the robustness to network modularity we need a kinetic description of the same level of description. Given a reaction system generated by the scheme above we assign a rate constant $\kappa_r$ to each reaction $r$ drawn from a normal distribution $N(\mu_{\text{rate}},\sigma_{\text{rate}})$ (the sign of $\mu_{\text{rate}}$ defines the direction of the reaction) and initial concentration $c_i$ of a substance $i$ in $N(\mu_{\text{conc}},\sigma_{\text{conc}})$ (setting negative concentrations to zero). From this starting point, we use the kinetic equation

$$\frac{dc_i}{dt}=\sum_r \kappa_r \pi_r(i)\prod_{j=1}^{S(r)} s_j^{\sigma_j} \qquad (2)$$

where the sum is over all reactions $r$ with $i$ as a product, where $\pi_{r(i)}$ is $i$'s multiplicity in the reaction $r$. To simulate the metabolic flux we also add source and sink terms to Eq. 2 for some metabolites. We let all the metabolites that are not substrates of any reaction be sinks (otherwise their mass would just accumulate) and all metabolites that are not a product of any reaction to be sources. In practice there will always be both sources and sinks in the generated reaction systems. (If the reaction systems would be generated in some other way one would need to put in sources and sinks explicitly.) We model the outflux by letting the sink-metabolites flow out of the system with a rate proportional to $\alpha$ times the concentration of the metabolite. In our simulations we use $\alpha=0.5$. We keep the inflow rate the same as the outflow rate so that the total mass is conserved. The inflow is distributed to the inflow metabolites in proportion to $\beta_i$, a random variable for each inflow metabolite drawn from a $N(\mu_{\text{in/out}},\sigma_{\text{in/out}})$ distribution when the reaction system is generated.

From the above setup, we run the system is until it converges (which it always does for the dynamic systems in question). We integrate the system with the Euler method (with time step $dt=10^{-5}$ until the time $t$ when

$$|c_i(t)-c_i(t')|<\varepsilon \text{ for all } \varepsilon \text{ and } t'\in[t,t+T] \qquad (3)$$

We use $\varepsilon=10^{-5}$ in this simulations. Higher precision in $dt$ or $\varepsilon$ does not change the outcome significantly. In this paper we use the parameter values $\mu_{\text{rate}}=0$, $\sigma_{\text{rate}}=1$, $\mu_{\text{conc}}=0$, $\sigma_{\text{conc}}=1$, $\mu_{\text{in/out}}=1$, $\sigma_{\text{in/out}}=1$ and $T=1$.

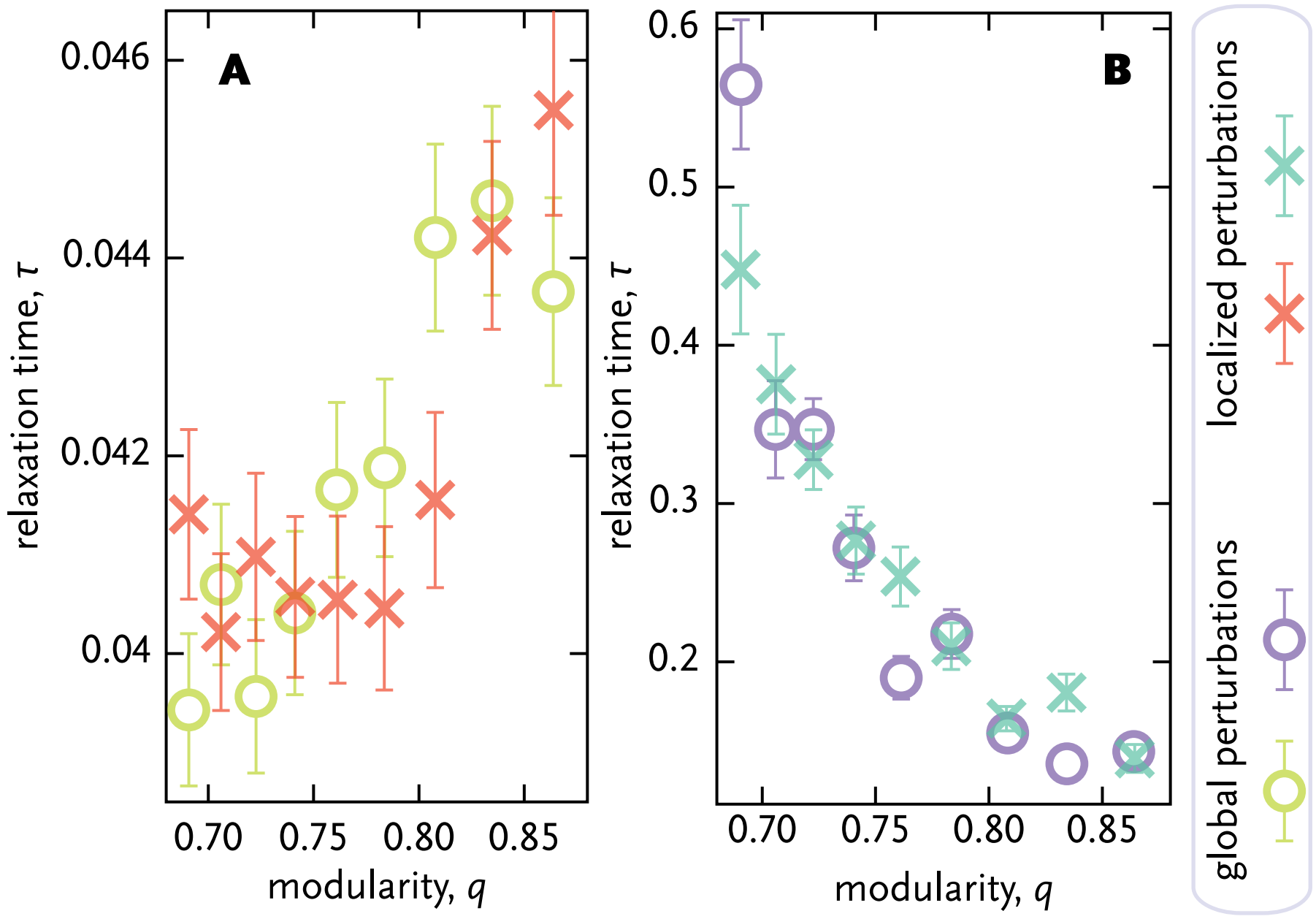

**Fig. 4. Relaxation time vs. modularity.** Panel A displays the corresponding data for perturbations in the concentrations of metabolites. Panel B shows the relaxation time for genetic perturbations within one module (circles) or the whole system (crosses). The data represents averages over more than 500 runs (the same runs as in Fig. 3.

### Genetic perturbations

Since we exclude genetic control and explicit enzymes in our reaction-system kinetics, we have to model the genetic perturbations indirectly. This is on the other hand quite straightforward. We replace $R_{\text{pert}}$ randomly chosen reactions following the same rules as when the reaction system was first constructed. For local perturbations, the reactions are chosen from one randomly selected cluster (identified by the cluster-detection algorithm above). A reaction is associated to the module to which a majority of its metabolites are categorized (if there is a tie, we select a cluster randomly). In this process, new metabolites will inevitably be generated and others possibly deleted. To conserve mass in case the number of metabolites changes, we split the mass of the deleted metabolites equally among the new. We also go over the system and update the sources and sinks in the same way as when the reaction system was constructed.

### Metabolic perturbations

Analogously to the genetic perturbations, we also require the metabolic perturbations to conserve the total mass. We control the magnitude of the perturbation by a parameter $\Xi$ by requiring that

$$\frac{\sum_{i\in\Omega}|\underline{m}_i-\overline{m}_i|}{\sum_{i\in\mathbf{M}}\overline{m}_i}=\Xi \qquad (4)$$

where $\overline{m}_i$ is the total mass of metabolite $i$ before the perturbation and $\underline{m}_i$ is the total mass after, and $\Omega$ is a set of metabolites. In practice the masses have a right-skewed, heavy tailed distribution (as observed in real systems (23)). This means that if we just continue adding metabolites randomly until the condition Eq. 4 is fulfilled, and $\Xi$ is not very small (we use $\Xi=5\%$), we will have to perturb a rather large fraction of the metabolites. To get around this problem, consider a set $\Upsilon$ of metabolite pairs. For the local perturbations, we choose a cluster (as detected by the algorithm above) at random as $\Omega$ and add pairs of metabolites picked at random to $\Upsilon$ until the condition is met or all there are no metabolites left in the cluster[1]. For the global perturbations we let $\Omega=\mathbf{M}$ and split the metabolites into two sets $\mathbf{M}_+$ and $\mathbf{M}_-$ where the total mass of any metabolite in $\mathbf{M}_+$ is larger than any metabolite in $\mathbf{M}_-$ and $\mathbf{M}_+$ is as small as possible such that the total mass of $\mathbf{M}_+$ is larger than

1. To facilitate the analysis, the model parameters need to be chosen so that this is a rare event.

2Ξ. In our simulations $\mathbf{M_-}$ always has more elements than $\mathbf{M_+}$. Then we add pairs where one metabolite is randomly selected from $\mathbf{M_+}$ and one is randomly selected from $\mathbf{M_-}$ until Eq. 4 is true.

### Robustness measures

Any measure of robustness should increase the more similar the system is before and after a perturbation. For biological functionality, it could be just as important to keep the concentrations of rare metabolites steady as those of the most abundant ones. Let $\bar{c}_i$ be the concentration of metabolite $i$ before the perturbation and $\underline{c}_i$ be the concentration after. A natural choice would be to take the average over the metabolites of the change $|\underline{c}_i - \bar{c}_i|$ rescaled by the typical concentration of $i$ as a measure of unrobustness (and thus its reciprocal value as a measure of robustness). As "typical concentration" one choice is the average. In practice, the metabolites that are very close to zero in concentration can give a rather large signal due just to numerical errors. To suppress such numerical noise, we rather use the quadratic mean, which decreases the expression's sensitivity to fluctuations in the denominator in the frequent situation that the concentrations are close to zero, thus putting a lower weight on the more uncertain terms. Our robustness measure thus becomes

$$r = \left( \frac{1}{|\Omega|} \sum_i \frac{|\underline{c}_i - \bar{c}_i|}{\sqrt{\underline{c}_i^2 + \bar{c}_i^2}} \right)^{-1} \quad (5)$$

where Ω is a set of metabolites and |...| denotes the absolute value of a number or the number of elements of a set. We consider two versions of this measure, one averaged over the whole set of metabolites, which we call system-wide perturbations $r$, and one averaged over the metabolites directly affected by the perturbations (the metabolites participating in a reaction catalyzed by a perturbed enzyme in the case of genetic perturbations or, trivially, the perturbed metabolites of a metabolic perturbation), which we refer to as focal robustness $r^*$.

### Acknowledgments

PH acknowledges support from the Swedish Foundation for Strategic Research, the Swedish Research Foundation and the WCU (World Class University) program through the National Research Foundation of Korea funded by the Ministry of Education, Science and Technology (R31–2008–000–10029–0).